\documentclass{article}
\usepackage{spconf,amsmath,graphicx}
\usepackage{booktabs}
\usepackage{multirow}
\usepackage{subfig}
\usepackage{todonotes}
\usepackage{hyperref}
\usepackage{enumitem}
\setlist{nosep, leftmargin=14pt}
\usepackage{hyperref} 
\hypersetup{
 colorlinks=true,
 linkcolor=blue,
 urlcolor=cyan,
 citecolor=black
}
\usepackage[capitalise,nameinlink]{cleveref}
\newcommand{\etal}{\textit{et~al.\ }}

\def\textBF#1{\sbox\CBox{#1}\resizebox{\wd\CBox}{\ht\CBox}{\textbf{#1}}}
\newsavebox\CBox

\title{A Stronger Baseline for Automatic Pfirrmann Grading of lumbar spine MRI using Deep Learning}

\name{\begin{tabular}{@{}c@{}}
Narasimharao Kowlagi\thanks{Corresponding Author: narasimharao.kowlagi@oulu.fi} \qquad 
Huy Hoang Nguyen \qquad 
Terence McSweeney\\ \vspace{-0.3cm} \\
Simo Saarakkala \qquad 
Juhani M\"a\"att\"a \qquad 
Jaro Karppinen \qquad 
Aleksei Tiulpin
\end{tabular}}
\address{University of Oulu}

\begin{document}
%
\maketitle
\begin{abstract}
This paper addresses the challenge of grading visual features in lumbar spine MRI using Deep Learning. Such a method is essential for the automatic quantification of structural changes in the spine,  which is valuable for understanding low back pain. Multiple recent studies investigated different architecture designs, and the most recent success has been attributed to the use of transformer architectures. In this work, we argue that with a well-tuned three-stage pipeline comprising semantic segmentation, localization, and classification, convolutional networks outperform the state-of-the-art approaches. We conducted an ablation study of the existing methods in a population cohort, and report performance generalization across various subgroups. Our code is publicly available to advance research on disc degeneration and low back pain.

\end{abstract}
\section{Introduction}
\label{sec:intro}

\begin{figure}[htbp]
    \subfloat[PG 2]{\includegraphics[width=0.23\linewidth]{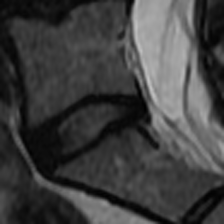}}\hfill
    \subfloat[PG 3]{\includegraphics[width=0.23\linewidth]{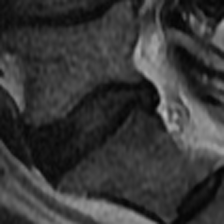}}\hfill
    \subfloat[PG 4]{\includegraphics[width=0.23\linewidth]{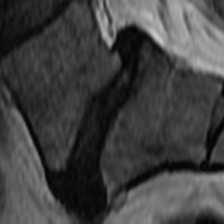}}\hfill
    \subfloat[PG 5]{\includegraphics[width=0.23\linewidth]{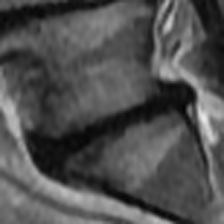}}\hfill
    \caption{Pfirrmann grades in L5-S1 as graded in the Northern Finland Birth Cohort dataset}
    \label{fig:pfirrmann_grades}    
    \vspace{-4mm}
\end{figure}

Lumbar Disc Degeneration (LDD) was found to be one of several reasons causing Low Back Pain~\cite{hartvigsen2018low}. Although there are several grading systems to quantify LDD, the Pfirrmann and Schneiderman scales are commonly used~\cite{samartzis2013disk}. Pfirrmann~\etal proposed a $5$-scale grading system using T2-weighted MR imaging~\cite{pfirrmann2001magnetic}. The grading system uses an algorithmic approach for LDD using signal intensity and disc homogeneity, demonstrated in~\cref{fig:pfirrmann_grades}. Over the past few years, several Deep Learning based methods were built to quantify LDD using MR imaging and the Pfirrmann grading system.

In a study by da Silva Barreiro~\etal~\cite{da2014semiautomatic}, the intervertebral discs (IVDs) were extracted using manually segmented binary masks, and an artificial neural network was used to perform Pfirrmann grading (PG). Later, SpineNet~\cite{JAMALUDIN201763} -- a framework that can automatically grade spine MRIs for different spine disorders, including LDD using the PG system was proposed. The authors used a graph-based approach to detect and group the vertebral bodies (VB) and discs in their work. Then, a VGG-M based architecture was used as the classifier. Newer architectures and methods were subsequently exploited in other works. For instance, SpineNetV2~\cite{windsor2022spinenetv2} used a UNet~\cite{ronneberger2015u} based architecture to improve vertebrae detection, and a ResNet~\cite{he2016deep} architecture was used for the classification of the spine disorders. Although the improvement was not substantial, this work highlighted the impact of design choices on the overall outcome. 

More recently, a transformer~\cite{dosovitskiy2020image} based architecture for spine grading~\cite{Windsor2022} was proposed and it outperformed the architecture designs proposed in the earlier works. In another related work, DeepSpine~\cite{lu2018deep} used a ResNeXt50~\cite{xie2017aggregated} architecture with the UNet model for spine segmentation to detect stenosis. Additionally, it is noteworthy to see an external validation of these methods~\cite{grob2022external}. 

\begin{figure*}[t!]
\centering
    \includegraphics[width=0.8\linewidth]{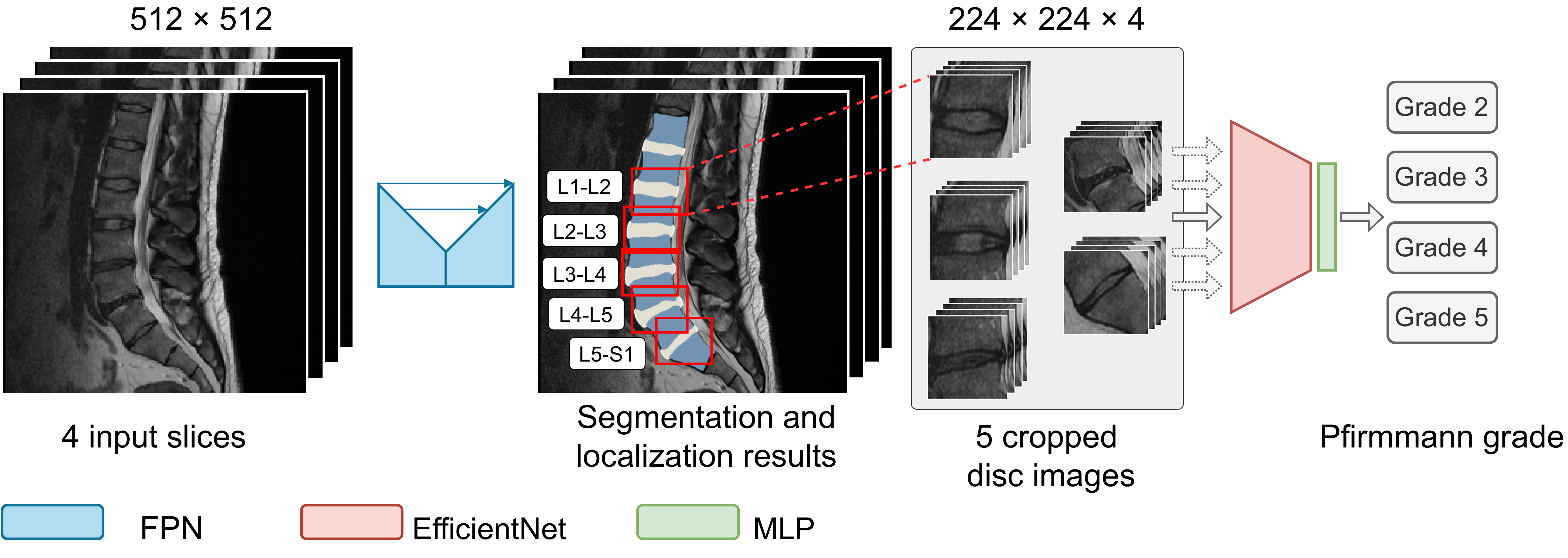}
    \caption{An end-to-end pipeline for quantifying LDD in the lumbar spine using Pfirrmann grading}
    \label{fig:pipeline}    
\end{figure*}

Although there is continuous improvement to create better spine grading models with newer architectures, we find a lack of thorough empirical evaluation in the existing research. Furthermore, current studies leveraged clinical data, but in reality, population cohorts are used in research. In this study, we take a step towards addressing these two challenges. Specifically, the contributions of our work are:
\begin{enumerate}
  \item We revisit the design choices used to create architectures for lumbar spine MRI grading, and propose a strong convolutional neural network (CNN) based method that outperforms the existing approaches in a systematic evaluation.
  \item Compare to the existing methods, we evaluate our methodology on a population cohort, which reflects the transferability of our results to downstream research applications.
  \item Finally, we make the code of this paper and the pre-trained models available to benefit further research.
\end{enumerate}

\section{Materials and Methods}

\label{sec:materialsandmethods}
\subsection{The Northern Finland Birth Cohort 1966}
The Northern Finland Birth Cohort (NFBC $1966$) dataset contains lumbar spine MRI data for $1500$ participants born in $1966$. Imaging data were acquired when the participants turned $46-47$ years old using $1.5$ Tesla GE Signa HDxt. Each slice has a resolution of $512\times512$ with a slice thickness of $4mm$ and a pixel spacing of $0.5mm$. 

For $1315$ subjects, the per-disc PGs were provided by a consensus of three experienced readers. Compared to the original definition, in the NFBC dataset, PGs 1 and 2 were treated as grade 2~\cite{takatalo2009prevalence}. In addition to the MR imaging, the participants from the same sub-group of NFBC were followed up with questionnaires to record information related to LBP. 

\subsection{Spine segmentation dataset}
A useful initial step for spine grading (\cref{fig:pipeline}), is to perform localization of functional spine units (FSUs)~\cite{Windsor2022} -- i.e. regions of interest (ROIs), containing two vertebral bodies and the associated disc. We approached this problem using semantic segmentation with $12$ classes: $6$ VBs, $5$ discs, and the background. 

An experienced clinician (T.M.), annotated a set of $250$ mid and para-sagittal slices with detailed boundaries for VBs and the discs ($512\times512$ pixels; original image resolution). Subsequently, we used an active learning approach by training UNet++~\cite{zhou2018unet++} with EfficientNetB3~\cite{tan2019efficientnet} to produce an initial segmentation model.

To extend the available data for model training, we used our initial model to produce predictions for unlabeled data, re-annotated poorly segmented samples, and included them in the training set. This process has been repeated several times to obtain segmentations for the whole dataset.

\subsection{Lumbar Spine segmentation}
An evaluation of segmentation architectures such as FPN~\cite{lin2017feature}, UNet~\cite{ronneberger2015u}, and UNet++~\cite{zhou2018unet++} was done through a series of ablation experiments that included encoder selection and hyperparameter tuning. 

We trained our models for $100$ epochs with an initial learning rate of $0.0001$ and a batch size of $10$ having an input shape of $512\times512$. The augmentation pipeline included random brightness, contrast, gamma, $\pm15$ degrees of rotation, and Gaussian blur. A weight decay of $0.0001$ was incorporated with the Adam optimizer~\cite{kingma2014adam}, and Jaccard loss~\cite{Iakubovskii:2019} was used for the optimization. 

We split the whole dataset into $2$ portions: $80\%$ for training and validation, and $20\%$ for independent testing. We then took $20\%$ of the former set for model selection. The mean Intersection over Union (mIoU) over independent runs based on $5$ random seeds was used to evaluate the model performance.

\subsection{Functional spine unit localization}
After finalizing the model architecture for segmentation, a post-processing step was incorporated to delineate $5$ FSUs from each slice of the lumbar spine MRI.
As such, we defined the FSU ROIs adaptively based on the bounding boxes inferred from the resulting segmentation mask. Subsequently, we rescaled these ROIs to $224\times 224$ pixels. Due to the volumetric nature of MRI, we obtained multiple slices with a shape of $224\times 224$ associated with each FSU. For efficiency, we utilized a subset of $n$ slices for PG.

\begin{figure*}[t!]

    \hspace*{\fill}
    \subfloat[Segmentation architectures]{\includegraphics[width=0.47\textwidth]{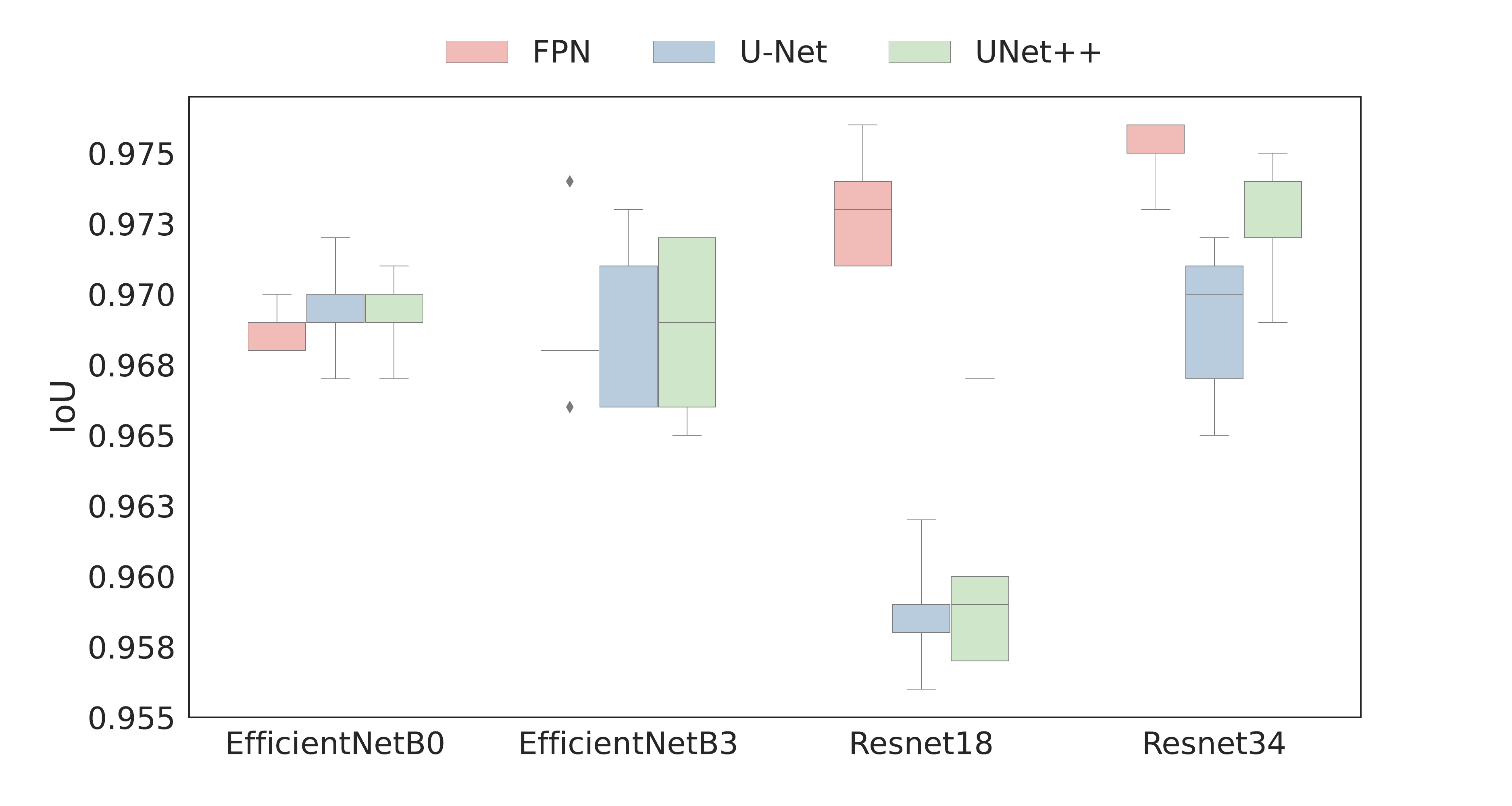}}\hfill
    \subfloat[Classification architectures]{\includegraphics[width=0.47\textwidth]{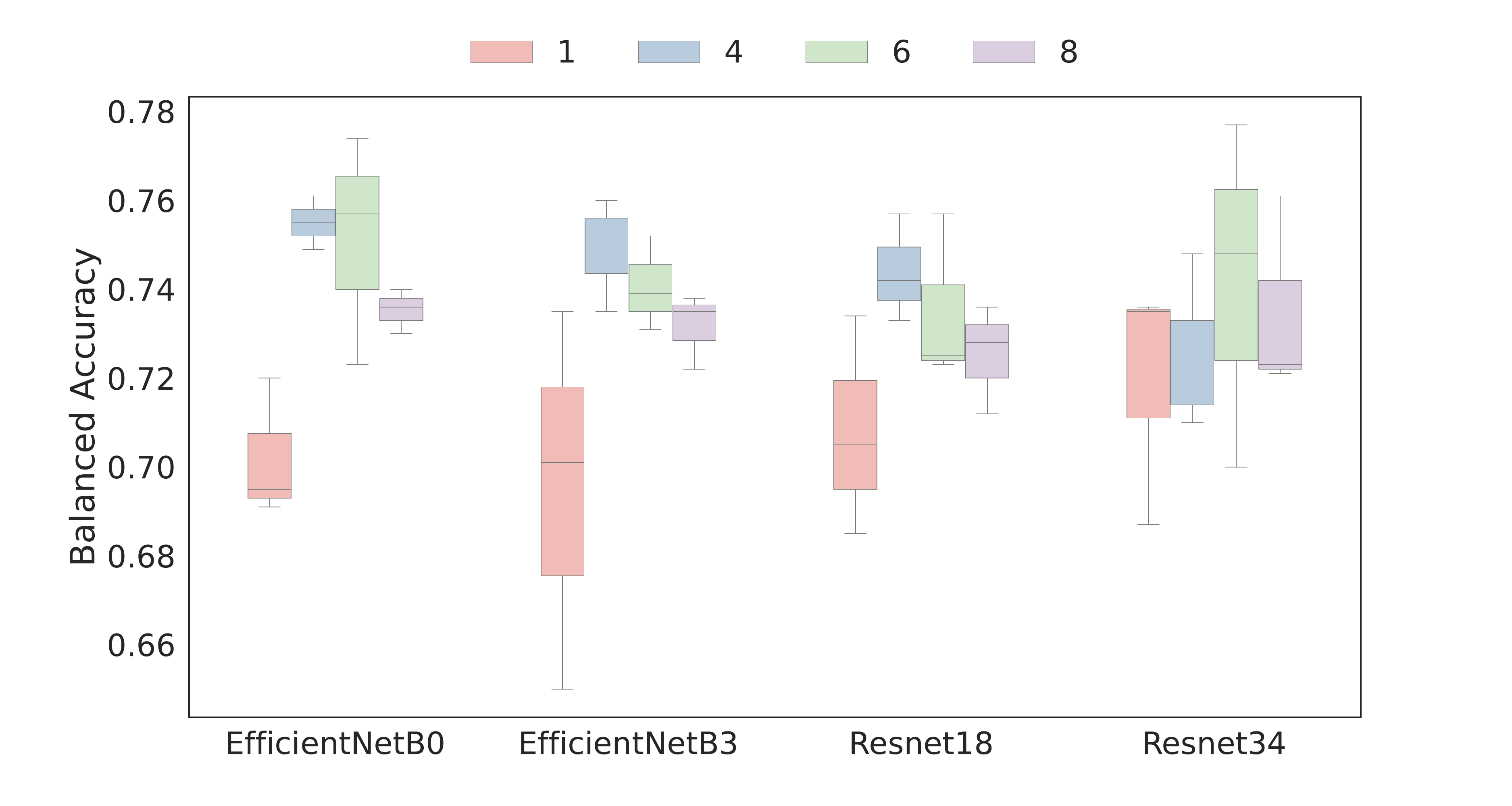}}
    \hspace*{\fill}
    \caption{A comparison of different architecture combinations on the segmentation and classification tasks. (b) $n\in\{1,4,6,8\}$ represents the number of slices.}
    \label{fig:seg_cls_arch}    
\end{figure*}

\subsection{Pfirrmann grading}
We utilized a common CNN to assess the LDD severity of the FSUs extracted in the previous step. As such, we built a DL architecture consisting of a feature extraction module for learning the representations of FSU slices, and a multi-layer perception (MLP) for PG. The candidates for the feature extractor were variants of ResNet~\cite{he2016deep} and EfficientNet~\cite{tan2019efficientnet}.
The MLP was a parametric non-linear function mapping the FSU representations to $4$ PGs.


We optimized the focal loss~\cite{Lin2017}, which is defined in the multi-class setting as  
\begin{equation}
    \mathcal{L}_{\mathrm{Focal}} = -\alpha_t(1-p_t)^\gamma \log p_t,
\end{equation}
where $p_t$ is the predicted probability of class $t$, and $\alpha$ and $\gamma$ are the loss hyperparameters. In our experiments, we used inverse class frequency to specify $\alpha$ and set  $\gamma=2$.

The data here were also split to have $20\%$ for testing. Similar to segmentation, we used $20\%$ of the training set for model selection. We trained all models using the Adam optimizer with a learning rate of $0.001$, a batch size of $20$, and a weight decay of $0.0001$. 

\section{Experiments}
\label{sec:experiments}

\subsection{Implementation details}
\subsubsection{Setup}
Through a series of ablation studies, we aimed to find the best models for each stage of our pipeline comprising vertebral segmentation, FSU localization, and spine grading. The pipeline was decoupled, and the training was performed independently for each stage. We implemented all the methods using the PyTorch framework, and conducted all the experiments on $2\ \times$ NVIDIA RTX A4000.


\subsubsection{Spine grading: Ablation study}
\label{ss:spinegrading}

To analyze the importance of different components, we conducted ablation studies over loss functions, samplers, schedulers, and slice volumes. To process the slice volumes and inputs, we replaced the first layer in the grading network with a randomly initialized one so that the number of input channels is equal to $n$.  

The base model was further trained with incremental settings such as consensus labels, ImageNet weights, augmentations, and Test Time Augmentations (TTA). At each incremental setting, the improvement is recorded. For augmentations, we used grid distortion \& downscaling in addition to the augmentations used for segmentation. We used a validation set to assess improvements.

\subsection{Results}
\subsubsection{Lumbar Spine segmentation}
A comparison of different architectures is shown in \cref{fig:seg_cls_arch}. From model comparisons, we found similar performance exhibited by UNet++ and FPN with the ResNet34 encoder. However, UNet++ suffered from high variance, making the FPN with ResNet34 the model of choice. FPN + ResNet34 combination scored an mIoU of $97.5\pm0.1$ over $5$ random seeds. 

\subsubsection{Spine grading model}
The base model for spine grading was chosen from the ablation study as discussed in \cref{ss:spinegrading}. The results of this can be seen in \cref{fig:seg_cls_arch}. Although EfficientNetB0 architecture with $6$ slices had the best Balanced Accuracy (BA), the standard error is higher than the model with $4$ slices. Therefore, the optimum number of slices to use was taken as $4$. Additionally, a more generalized observation shows that the model performance deteriorates as more slices are used. \cref{tbl:additional_tricks} shows the incremental improvement of the base model over a series of critical settings that contributed to a significant boost in the model's performance. 

\begin{table}[ht!]
\caption{Incremental improvement in model performance with different design choices}
\resizebox{\linewidth}{!}{
\begin{tabular}{lcccc} \toprule\
Setting & \multicolumn{1}{l}{Balanced Accuracy} & \multicolumn{1}{l}{Cohen's Kappa} & \multicolumn{1}{l}{Lin's CCC} \\ \midrule
Base Model             & $75.5\pm0.3$& $69.0\pm0.6$& $85.1\pm0.3$\\
With Consensus Grading & $77.6\pm0.3$& $69.6\pm0.3$& $85.8\pm0.4$\\
With Imagenet weights  & $78.8\pm1.3$& $70.7\pm0.6$& $86.8\pm0.5$\\
With Augmentations     & $81.1\pm0.8$& $72.6\pm0.6$& $88.1\pm0.4$\\
With TTA               & $\mathbf{81.7\pm0.6}$&$\mathbf{73.8\pm0.4}$&$\mathbf{88.5\pm0.1}$\\
\bottomrule
\end{tabular}     
}
\label{tbl:additional_tricks}
\end{table}

\subsubsection{Spine grading: Comparison to baselines}
\begin{figure}[ht!]
    \includegraphics[width=1\linewidth]{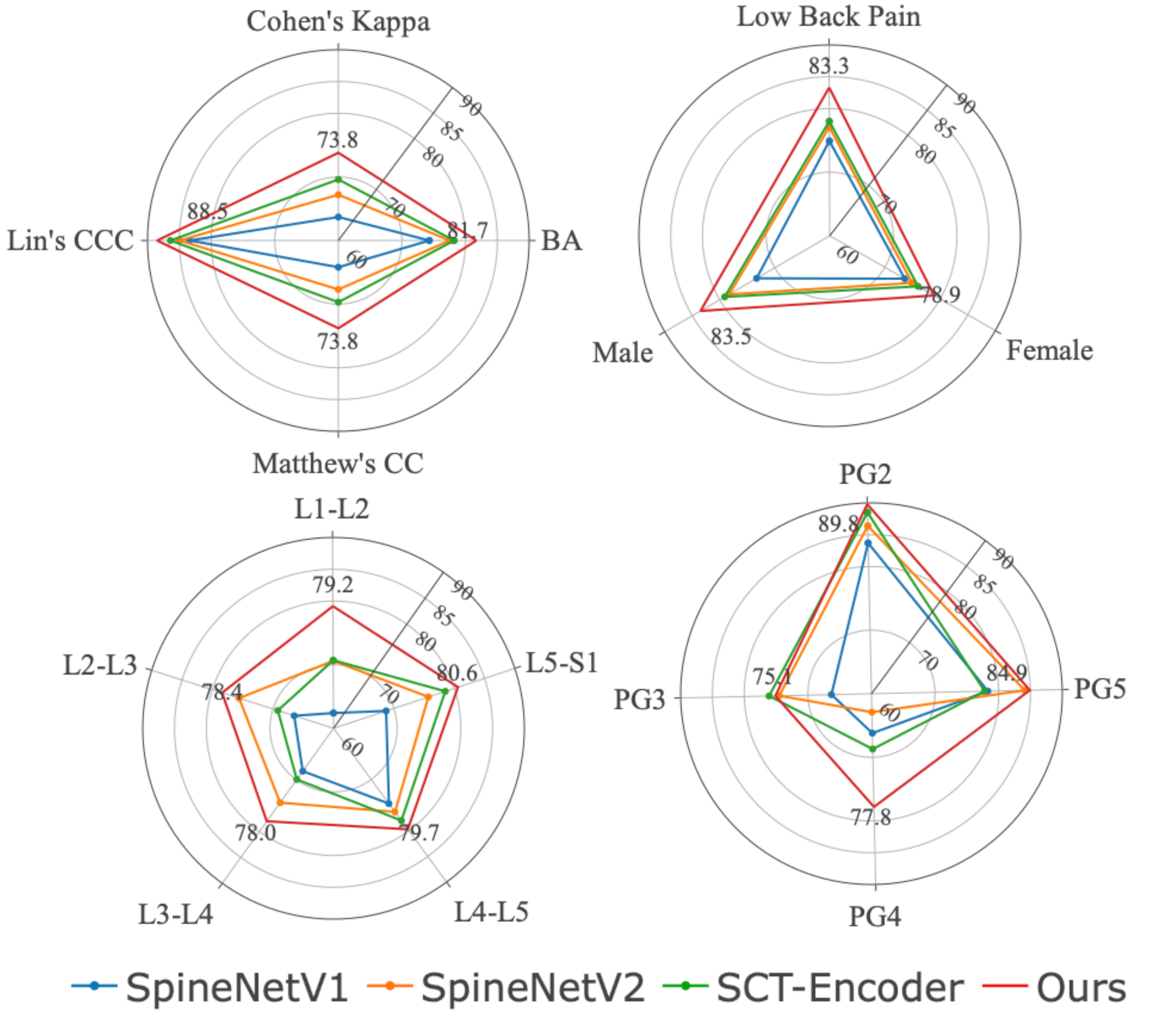}
    \caption{Inter-rater reliability and BA of our model compared to other methods at more granular levels of Sex, IVD, and PGs}
    \label{fig:radar_plot}    
\end{figure}

We compared our work with the current state-of-the-art methods that quantified LDD using PG. To ensure a fair comparison with our work, all the models are trained on the NFBC dataset with the weights initialization, hyperparameters, and loss functions prescribed by the authors in their work. While SpineNetV1 and SpineNetV2 are trained with their proposed architectures, we modified Context-Aware Transformers for our baseline by implementing the encoder part of the original network. This is due to the limitation on data availability for MRI sequences used in their method. A comparison of these methods with our work is shown in \cref{tbl:spinenetv2_compare}. Additionally, we extended the comparison beyond BA to measure the model performance at different subgroups as shown in \cref{fig:radar_plot}. Compared to our work, the prior works do not generalize well on our dataset when assessed on IVD and PG levels.

\begin{table}[ht]
 \centering
 \caption{Benchmark performance of our model against state-of-the-art spine grading models}
 \label{tbl:spinenetv2_compare}
 \resizebox{\linewidth}{!}{
 \begin{tabular}{lccccc} \toprule\
 Model Architecture  & \multicolumn{1}{l}{Balanced Acc} & \multicolumn{1}{l}{Cohen's Kappa} & \multicolumn{1}{l}{Lin's CCC}\\ \midrule
 SpineNetV1 & $74.3\pm0.6$ & $63.7\pm1.2$ & $83.4\pm0.7$\\
 SpineNetV2 & $77.5\pm0.5$ & $67.2\pm0.9$ & $85.0\pm0.6$\\
 SCT-Encoder& $78.2\pm0.8$ & $69.6\pm0.9$ & $86.4\pm0.5$\\
 \textBF{Ours}& $\mathbf{81.7\pm0.6}$&$\mathbf{73.8\pm0.4}$&$\mathbf{88.5\pm0.1}$\\
 \bottomrule
 \end{tabular}     
 }

 \end{table}


\section{Discussion}
\label{sec:discussion}
In this work, we revisited different architectures used for LDD assessment using the PG system and proposed a robust baseline that outperformed existing methods. An emphasis is placed on the importance of design choices and settings contributing to improved model performance. Our model exhibits more generalized characteristics and does not bias towards a specific subgroup, as observed in other methods. We believe that our study will contribute towards better robust stratification of patients for LDD using lumbar spine MRIs. Our code and pre-trained models are released at~\url{https://oulu-imeds.github.io/ISBI2023_PfirrmannGrading/} to advance further studies.

\section{Compliance with ethical standards}
\label{sec:ethics}
 NFBC1966  participants  gave  their  informed  consent  and  the  Regional  Ethics  Committee  of  the  Northern Ostrobothnia  District approved the study.
\section{Acknowledgements}
\label{sec:acknowledgements}
This work was funded by the Marie Sklodowska-Curie Actions (MCSA), International Training Network, under grant agreement 955735 for the Disc4All consortium. This work was also supported by funding from the
Academy of Finland (Profi6 336449 funding program).

\bibliographystyle{IEEEbib}
\bibliography{strings,refs}

\end{document}